\documentclass[]{elsart}
\usepackage{epsfig}
\usepackage{amssymb}
\usepackage{amsmath}
\usepackage{longtable}

\newcommand{\one}{\ensuremath{1\!{\rm l}}}

\newcommand{\rmd}{\ensuremath{{\rm d}}}
\newcommand{\bs}[1]{\ensuremath{\boldsymbol{#1}}} % mathe. vektor
\begin{document}

\begin{frontmatter}

\title{The $A_y$-Problem in Refined Resonating Group Model calculations 
  for ($p-^3{\rm He}$) Scattering}
\author[erlangen,victoria]{C. Rei{\ss}}
\ead{reissc@kph.uni-mainz.de}
and
\author[erlangen]{H.M. Hofmann}

\address[erlangen]{
  Institut f{\"u}r Theoretische Physik III, 
  Universit{\"a}t Erlangen-N{\"u}rnberg,
  Staudtstra{\ss}e 7, D-91058 Erlangen, Germany
}

\address[victoria]{
  Department of Physics and Astronomy,
  University of Victoria,
  3800 Finnerty Rd., Elliot Bldg.,
  Victoria, BC V8P 1A1, Canada
}

\begin{abstract}
  We report on a microscopic Refined Resonating Group Model (RRGM) 
  calculation of scattering of $p$ off ${}^3{\rm He}$
  employing the Argonne-$v_{14}$ and the Bonn nucleon-nucleon potentials
  without three-nucleon forces at low energies up to 30\,MeV. 
  The calculated phase-shifts
  verify the well-known proton analyzing power $A_y$-problem. We demonstrate
  that with corrected ${}^3P_2$ phase-shifts  experimental
  differential cross-section and analyzing power data can be explained.
\end{abstract}

\begin{keyword}
phase-shift \sep analyzing power \sep Refined Resonating Group Model \sep
realistic $NN$-interactions
\PACS  21.45.+v \sep 24.70.+s \sep 25.40.Cm
\end{keyword}
\end{frontmatter}

%##############################################################################
% INDRODUCTION 
%##############################################################################

\section{Introduction}

It is widely known that realistic nucleon-nucleon ($NN$) forces cannot
reproduce the $\rm ^3H$ and $\rm ^3$He binding energies. 
The $A_{\rm y}$ analyzing power in $(N-d)$ scattering is
still not reproduced by adding three-nucleon interactions \cite{AY}.  
The 30\% deviation of $A_{\rm y}$ can be
resolved by tiny changes  on the order of $0.1^\circ$ in the $(N-d)$ 
scattering phase-shifts
\cite{PD_3MEV,KIEVSKY_ND,nd_psa}.
It was shown in \cite{HE4} that although a realistic $NN$-force can generally
reproduce the $^4$He system, there remain differences,
most notably in the analyzing powers.  
Recently Fonseca et al. \cite{fonseca02} studied 
the $(n-{}^3{\rm He})$ scattering system using an AGS-approach.
They calculated the cross-sections and vector analyzing 
powers using the $NN$-interactions Bonn-CD \cite{bonncd}, Nijmegen II
\cite{nijmegen2} and Bonn-B \cite{bonnb} without $3N$-forces 
and pointed out that the $(n-{}^3{\rm He})$ system
may turn into a $A_y$ puzzle like it is known in $(n-d)$ scattering
and may be cured by $3N$-forces.
The detailed  studied ${}^4{\rm He}$ system \cite{TILLEY_A4} 
is unfortunately extremely difficult to
describe due to the $\rm^4He$ bound-state and particularly 
the complex resonance-structure.
For that reason we investigate the much simpler system $(p-{}^3{\rm He})$,
the ${}^4{\rm Li}$. 
In the studied energy range up to 30\,MeV is a lot of data available.
And we have restricted ourselves on realistic $NN$-interactions
without three particle forces 
namely the Argonne-$v_{14}$-potential
\cite{Wir84} and the Bonn-potential in the form of Ref. \cite{Kel89}.

The paper is organized as follows: In section \ref{sec:rrgm} 
we give a brief introduction to the RRGM.
We present in section \ref{sec:model} for each of the interactions 
(Argonne-$v_{14}$ and Bonn) two model-spaces (a medium and large one) 
for the used RRGM scattering calculation. 
Then we compare and discuss in section \ref{sec:results}
the obtained phase-shifts $\delta$, analyzing powers $A_y$
and differential cross-section $\rmd\sigma/\rmd\Omega_{\rm c.m.}$
with experimental data and  demonstrate the origin of 
the $A_y$-problem.

\section{Refined Resonating Group Model with Distortion channels}
\label{sec:rrgm}

%##############################################################################
% RRGM
%##############################################################################
\subsection{Refined Resonating Group Model}

This calculation is based on the Refined Resonating Group Model (RRGM) 
\cite{Hof87} 
with distortion channels \cite{Wol87}. Therefore 
we briefly summarize the RRGM in the following.

The model employed is restricted to a two fragment description. That means
the model is not able to describe the three particle breakup reactions.
Therefore a scattering channel consists of two fragments $i=1,2$ with total
angular momentum $J_{i}$ which couple to the channel spin $S_{\rm c}$.
The channel spin $S_{\rm c}$ couples with the 
relative orbital angular momentum $L_{\rm rel}$ 
to the total angular momentum $J$ of the channel 
${}^{2S_{\rm c}+1}{L_{\rm rel}}_{J}$.
The binding energy and the wavefunction of one fragment is obtained
using the variational principle of Ritz. The solution of the
Schr{\"o}dinger-Equation for the scattering problem is won by
the variational principle of Kohn-Hulth{\'e}n \cite{Koh48}.

The Hamilton-operator $H$ for a $N$-particle system with
a two-body-force is given by:
\begin{equation}
  H(1,...,N) = \sum^{N}_{i=1} T_{i} + \sum^{N}_{i,j=1 \atop i<j } V_{ij}
  \quad. 
\end{equation}
Using momentum conservation, the center-of-mass energy $T_{\rm{c.m.}}$ 
is separated off.
The restriction to
a two-fragment model allows to formulate a translationally invariant
Hamilton-operator $H'$ consisting of fragment and relative-motion parts.
The potential term 
becomes shortranged  by subtracting
the Coulomb-potential $Z_{1}Z_{2}\,e^{2}/R$ with the
relative coordinate $\bs{R}$ between the fragments:
\begin{eqnarray}
  H'(1,...,N) & = & H_{1}(1,...,N_{1}) + H_{2}(N_{1}+1,...,N) + T_{\rm{rel}} +
  Z_{1}Z_{2}\,e^{2}/R\nonumber\\ 
  \nonumber\\ 
  && +  \left( \sum_{i \in \{1,...,N_{1}\} \atop j \in \{N_{1}+1,...,N\}}
    V_{ij} - Z_{1}Z_{2}e^{2}/R\right)\quad. 
\end{eqnarray}
With the variational principle of Kohn--Hulth{\'e}n \cite{Koh48}
\begin{equation}
  \delta \left(\langle \Psi_{l}\left| H' - E \right|\Psi_{l}\rangle -
    \frac{1}{2} a_{ll}\right) = 0\label{var}
\end{equation}
we determine the solution of the Schr{\"o}dinger-Equation
where the scattering wavefunction is denoted by $\Psi_{l}$.
From the reaction matrix $a$ the scattering matrix $S$ is calculated
via the Cayley--Transformation:
\begin{equation}
  S = \left( \one + i a \right)\left( \one - i a \right)^{-1}
  \quad.\label{cayley}
\end{equation}
The index $l$ in Eq.(\ref{var}) is the label of the 
corresponding boundary condition.
The ansatz for the wavefunction $\Psi_{l}$ is
\begin{equation}
  \Psi_{l} = {\mathcal A} \left\{ \sum_{k=1}^{n_{k}} \Psi_{\rm{c}}^{k} \cdot
    \Psi_{\rm{rel}}^{lk}\right\}
  \quad,\label{psil}
\end{equation}
with the antisymmetrization operator 
\begin{equation}\label{antisym}
  {\mathcal A}=\sum_{P}(-1)^{P}P
\end{equation}
where $P$ is the permutation over all particles, 
$n_{k}$ is the number of channels, $\Psi_{\rm{c}}^{l}$
is the channel function and  $\Psi_{\rm{rel}}^{lk}$ is
the relative wavefunction:
\begin{equation}
  \Psi_{\rm{rel}}^{lk}(R) = \delta_{kl} {\rm F}_{k}(R) + a_{lk} {\widetilde
    {\rm G}}_{k}(R) +
  \sum_{m} b_{lkm} \chi_{km}(R)
  \quad.
\end{equation}
${\rm F}_{k}$ and ${\widetilde {\rm G}}_{k}$ are regular and  regularized
irregular Coulomb-functions for the correct description of the asymptotic behavior
of the wavefunction. $\delta_{kl}$ is here the  Kronecker-Symbol. 
The  $\chi_{km}$ are square-integrable functions for the description
of the wavefunction in the interaction region. The  variation of the  $a_{lk}$ 
and the $b_{lkm}$ yields with Eq.(\ref{cayley}) the $S$-Matrix:
\begin{equation}
S_{kl}=\eta_{kl}\ e^{2i\delta_{kl}}
\end{equation}
with phase-shifts $\delta_{kl}$ and channel coupling-strengths $\eta_{kl}$.

The scattering wavefunction $\Psi_{l}$ (see Eq.(\ref{psil})) 
was already explained but without the channel part. We look only at
one term in Eq.(\ref{psil}). It is convenient to make the following
ansatz for the channel function $\Psi_{\rm{c}}$: 
\begin{equation}
  \Psi_{\rm{c}} 
  = 
  \left[ 
    \frac{{\rm Y}_{L_{\rm rel}}
      (\hat{\bs{r}})}{r} \otimes \left[  \phi_{1}^{J_{1}} \otimes
      \phi_{2}^{J_{2}} 
    \right]^{S_{\rm{c}}}
  \right]^{J}
  \quad.\label{eqn:channelfunction}
\end{equation}
We use $\phi_{i}^{J_{i}} (i = 1,2)$ for the translational invariant
wavefunction of the $i$th fragment with spin $J_{i}$. 
In Eq.(\ref{eqn:channelfunction})
square brackets indicate the angular momentum coupling.
In the case of scattering calculation 
the coordinate $\bs{r}$ is the relative coordinate  $\bs{R}$. 
If $\Psi_{\rm{rel}}(\bs{r})$ is a bound-state wavefunction
then there will be no coupling to the channel spin  $S_{\rm c}$
and the coordinate $\bs{r}$ will be the Jacobi-coordinate between
the center-of-mass of the both fragments.
The  fragment function $\phi^{J}$ is build of the spin-isospin function  
$\Xi$ and of the  spatial function  $\chi$:
\begin{equation}
  \phi^{J} 
  = 
  \sum_{l_{I},S,\alpha} 
  \left[ 
    C_{\alpha}^{l_{I}LS} \chi_{\alpha}^{l_{I}} \Xi^{ST}
  \right]
\quad.
\end{equation}
The set of inner orbital angular momenta in the fragments is labeled with $l_{I}$. 
The corresponding
fragment consists of $n$ nucleons. The spin-isospin functions 
$\Xi$ are coupled to good total spin and good isospin. 
The Clebsch-Gordan-coefficients 
of the coupling of the orbital angular momenta $l_{1}, l_{2},..., l_{n-1}$ to the
total orbital angular momentum $L$, the coefficients of the coupling of the spins 
and isospin
are coupled to $S$ and $T$ and the coefficients of the super-position of the
radial dependencies $\alpha$ are expressed by the factor $C_{\alpha}^{l_{I}LS}$.
The $\chi_{\alpha}^{l_{I}}$ are the square-integrable spatial functions.
They consist of nucleon relative functions $\chi_{\alpha,k}^{l_{k}}$:
\begin{equation}
  \chi_{\alpha}^{l_{I}} 
  =   
  \prod_{k=1}^{n-1}
    \chi_{\alpha,k}^{l_{k}}
    \quad.
\end{equation}
The  nucleon relative function consists
from a Gaussian function with width parameter $\beta_{\alpha,k}$ and 
from a Solid-Spherical-Harmonics $\mathcal Y$ \cite{Edm74}:
\begin{equation}
  \chi_{\alpha,k}^{l_{k}} 
  = 
  \exp \left\{ - \beta_{\alpha,k}\,\rho_{k}^{2}
  \right\} {\mathcal Y}_{l_{k},m_{k}}(\bs{\rho}_{k}),
\end{equation}
where $\bs{\rho}_{k}$ is the Jacobi-coordinate between the 
($k+1$)th nucleon and the center-of-mass of the  nucleons ($1,2,...k$).

All spatial functions are in this model finally parametrized 
with Gaussian functions.

\subsection{Distortion Channels}

Distortion channels in the RRGM are unphysical bound-states which
can be chosen arbitrarily as far as they are linear independent and as long as
they have right total angular momentum and total parity. 
They are allowed to violate e.g. the channel spin $S_{\rm c}$ of the scattering
channels. Distortion channels have the asymptotics of bound-state channels. The idea
of including distortion channels is to enlarge the model-space in the region of
interaction. Therefore distortion channels can be used to describe physical 
structures which can not be described by the RRGM or are not included in
the scattering channels e.g. deformations of the fragments, excited
fragments and scattering-states which are not yet opened channels.
A more detailed description of the approach using distortion channels with
the RRGM can be found by R. W{\"o}lker \cite{Wol87}.

%##############################################################################
% MODEL SPACES
%##############################################################################
\section{Model-Spaces for $(p-{}^3{\rm He})$ Scattering}\label{sec:model}

\begin{figure}[h]
  \epsfig{file=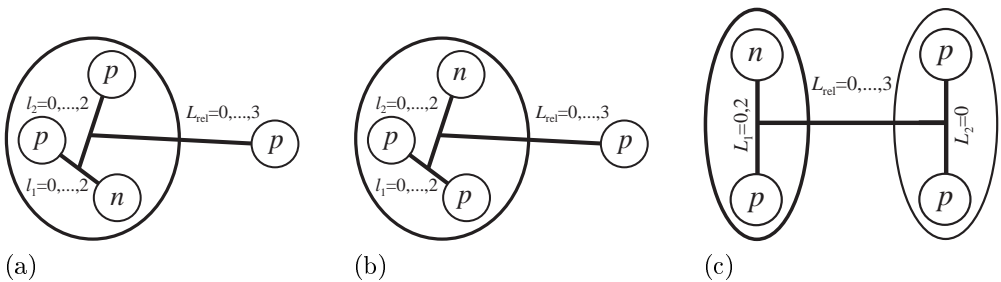}
  \caption{Panel (a) and (b) show the  $(p-{}^{3}{\rm He})$--structures 
    and subfigure (c) shows the $(d-{}^{2}{\rm He})$--structure used 
    to describe the $(p-{}^{3}{\rm He})$--scattering-system
    with $L_{\rm rel}\leq 3$.}\label{fig:struct}
\end{figure}

For the description of the 
${}^{4}{\rm Li}$-scattering-system with 
$L_{\rm rel}\leq 3$  and $J^\pi=0^\pm,\,1^\pm,\,2^\pm,\,3^+$
we have used the sets of Jacobi-coordinates with 
orbital angular momenta as indicated in Fig.\ref{fig:struct}.
We have studied this scattering-system with the Argonne-$v_{14}$- ($av_{14}$) 
and the Bonn-potential. 
For the ${}^{3}{\rm He}$-subsystem we have used two different model spaces:
a medium system (labeled with HeM) with 5 basis-vectors 
and  a large   ${}^{3}{\rm He}$-subsystem (labeled with HeG) with 23 basis-vectors.
The $(p-{}^{3}{\rm He})$-scattering-system using the large
${}^{3}{\rm He}$-subsystem was studied without $F$-wave
scattering because the $F$-wave phase-shifts turned out  very small 
in the medium system $(p-{}^{3}{\rm He})$-system.
Table \ref{tab:struct} shows the used 
model-spaces in compact form. The sets of width parameters consisting 
of $n_1$ elements on the first Jacobi-coordinate and $n_2$ on the second
indicated in brackets are assigned to spin-isospin-functions
with the choice of angular momenta. 
The sets of widths are determined by a non-linear optimization
\cite{Winkler97}.
In the scattering calculation for the large  ${}^{3}{\rm He}$-subsystem
the smallest width on the second coordinate was dropped due
to the fact that there was practically no improvement of the ${}^{3}{\rm He}$
wave-function.
For the rearrangement channel $(d-{}^2{\rm He})$ we used
for the description of the deuteron $d$ three width parameters for the $S$-wave
and two for the $D$-wave. The unbound ${}^2{\rm He}$-system
is described with an $S$-wave using the $S$-wave width set of
the deuteron. 
The rearrangement channel $(d-{}^{2}{\rm He})$ opens in the medium system 
at $5.7$ MeV using the Argonne-$v_{14}$- and at $6.5$ MeV using the
Bonn-interaction. For the large model these energies are $7.1$ MeV ($av_{14}$)
and $7.6$ MeV (Bonn).
The binding-energies $E_{\rm B}$, the charge ${\rm r.m.s.}$-radii and the 
$D$-state probabilities of the models are summarized in Tab. \ref{tab:binding} 
and Tab. \ref{tab:rms}. 
As expected the $D$-state probabilities 
for the  Argonne-$v_{14}$-interaction  are appreciably higher than those for 
the Bonn-potential, see table \ref{tab:rms}.
With increasing the model-space the 
binding-energy is improved and also the ${\rm r.m.s.}$-radii. 
The ${\rm r.m.s.}$-radii are smaller than the experimental ones and 
they are even smaller for smaller model-spaces.
Naturally one would expect for models which 
underbind the system to generate larger ${\rm r.m.s.}$-radii.
Here, however, in the smaller model-spaces those width parameters,
which generate the larger radii, are missing.
A basis-vector of the ${}^{4}{\rm Li}$-scattering-system is
therefore a  spin-isospin function consisting of one neutron and three protons
with a fixed orbital angular momentum configuration and the appropriate set
of width parameters. The set of distortion channels for both models was obtained
by using  the $(d-{}^{2}{\rm He}^*)$ and $(d-{}^{2}{\rm He}^{**})$ channels
as distortion-channels because they open at energies higher than $30$ MeV
and are therefore out of the studied range of energies. 
In addition as much as possible
of the existing basis-vectors have been used to form the maximum of 
linear independent distortion-channels. Every additional distortion 
channel contains therefore only  one basis-vector. 
For the scattering-channel the set of widths ${{\rm w}_{\rm rel}}$
was used on the relative coordinate, where for the distortion channels only the subset
${{\rm w}_{\rm dist}}$ was taken on that coordinate.

\begin{table}[ht]\begin{longtable}{l|llllll}
\hline\hline\hline
Model & $[[{\rm T}_1{\rm T}_2]^{s}{\rm T}_3]]^{S}$ 
& \multicolumn{5}{c}{$(l_1l_2)^{L}$}\\
\hline
HeM
\hfill(3+2)
& $[[np]^{0,1}p]^{1/2}$ & $(00)^{0}$ \\
& $[[pp]^{0}n]^{1/2}$   & $(00)^{0}$ \\
& $[[np]^{1}p]^{3/2}$ & $(20)^{2}$ & $(02)^{2}$ \\
\hline\hline 
HeG
\hfill(3+3) 
& $[[np]^{0,1}p]^{1/2}$ & $(00)^{0}$ & $(11)^{0,1}$ \\
& $[[pp]^{0}n]^{1/2}$ & $(00)^{0}$  \\
& $[[pp]^{1}n]^{1/2}$ & $(11)^{0,1}$ \\
& $[[Np]^{1}\bar{N}]^{3/2}$ & $(11)^{1,2}$ \\
& $[[np]^{0,1}p]^{1/2}$ & $(22)^{0,1}$ \\
& $[[pp]^{0}n]^{1/2}$ & $(22)^{0,1}$ \\
& $[[np]^{1}p]^{3/2}$ & $(20)^{2}$ & $(02)^{2}$ & $(22)^{1,2}$ \\
\hline\hline\hline
 ${\rm w}_{\rm rel}\,\,[{\rm fm}^{-2}]$ \hfill(20) 
 & 12.95665%0 
 & 5.134670%0 
 & $\underline{2.947287}$ %0 
 & 1.342339%0 
 & $\underline{0.821446}$%56 
\\
 $\underline{{\rm w}_{\rm dist}\,\,[{\rm fm}^{-2}]}$ \hfill(5) 
 & 0.444741%3 
 & $\underline{0.293900}$ %0 
 & 0.169016%75 
 & $\underline{0.118524}$ %36 
 & 0.084300%0 
\\
 & $\underline{0.050011}$ %0 
 & 0.025737%69 
 & 0.013852%0 
 & 0.007143%29 
 & 0.003852%19 
\\
 & 0.001857%3 
 & 0.000973%26 
 & 0.000562%19 
 & 0.000277%65 
 & 0.000101%0 
\\
\hline\hline\hline
\caption{The model-spaces for the ${}^3{\rm He}$-system 
  is given in the upper two third of the table.  
  The sets of widths are assigned to the spin-isospin-function
  and angular momentum structures.
  The number of sets of widths on the first and the second
  Jacobi-coordinates is indicated in brackets.
  ($N$ is $p$ or either $n$ and therefore $\bar{N}\not =N$ denotes $n$
  or $p$). 
  The sets of widths for the relative coordinate in the 
  ${}^4{\rm Li}$-scattering-system for the scattering- ${\rm w}_{\rm rel\ }$
  and distortion-channels ${\rm w}_{\rm dist}$ (underlined) 
  are given in the lower third of the table.
  }\label{tab:struct}
\end{longtable} \end{table}

\begin{table}[ht]
\begin{longtable}{lr|rr|rr}
\hline\hline
\multicolumn{2}{c|}{$NN$-Interaction} &
\multicolumn{2}{|c|}{Argonne-$v_{14}$} &
\multicolumn{2}{|c}{Bonn}\\
\hline
$E_{\rm B}^{}(d)$  & [MeV] 
& $-$1.878  %$15.6 \%$ 
& 
& $-$1.911  %$14.1 \%$ 
&\\
$E_{\rm B}^{\rm max}(d)$ & [MeV] 
& $-$2.222  %$0.00 \%$  -2.225
& & $-$2.222  %$0.14 \%$ 
&\cite{Kel89} \\
\cline{3-6}
$E_{\rm B}^{\rm exp}(d)$ & [MeV] & \multicolumn{3}{|c}{$-$2.225}  &\cite{Audi95}\\ 
\hline
$E_{\rm  B}^{\rm M}({}^{3}{\rm He}^{1/2+})$ & [MeV] & $-$5.588  %$27.6 \%$ 
& & 
$-$6.287  %$18.5 \%$ 
& \\
$E_{\rm  B}^{\rm G}({}^{3}{\rm He}^{1/2+})$ & [MeV] & $-$6.895  %$10.7 \%$ 
& &
$-$7.416  %$3.91 \%$ 
& \\
$E_{\rm B}^{\rm max}({}^{3}{\rm He}^{1/2+})$ & [MeV]  &
$-$6.949 %$9.96 \%$ 
& & $-$7.492  %$2.93 \%$
& \\
\cline{3-6}
$E_{\rm B}^{\rm exp}({}^{3}{\rm He}^{1/2+})$ & [MeV] & 
\multicolumn{3}{|c}{$-$7.718} & \cite{Til87}\\ 
\hline\hline
\caption{
  The binding-energies $E_{\rm B}$  
  of the used model-spaces for the
  fragments in the ${}^4{\rm Li}$-scattering-system are summarized.
  For comparison the experimental data (exp) and the best result of 
  a RRGM calculation (max) are shown for the different studied
  $NN$-interactions Argonne-$v_{14}$ and Bonn.
}\label{tab:binding}
\end{longtable}
\end{table}

\begin{table}[ht]
\begin{longtable}{lr|rrl|rrl}
\hline\hline
\multicolumn{2}{c|}{$NN$-Interaction} &
\multicolumn{3}{|c|}{Argonne-$v_{14}$} &
\multicolumn{3}{|c}{Bonn}\\ 
\hline
$r_{\rm r.m.s.}^{}(d)$ & [fm] 
&   1.80   & (5.37\%)   %$7.69 \%$ 
& & 1.77   & (4.51\%)   %$9.23 \%$ 
&   \\
$r_{\rm r.m.s.}^{\rm max}(d)$ & [fm] 
&   1.98  & (6.08\%)   %$-1.54 \%$ 
& & 1.98  & (4.78\%)   %$-1.54 \%$ 
& \\%\cite{Kel89}  \\
\cline{3-8}
$r_{\rm r.m.s.}^{\rm exp}(d)$        & [fm] & 
\multicolumn{5}{|c}{2.13} & \cite{sick98} \\ 
\hline
$r_{\rm r.m.s.}^{\rm M}({}^{3}{\rm He}^{1/2+})$ & [fm] & 
      1.72  & (7.47\%) %$15.1 \%$ 
&   & 1.62  & (5.93\%) %$20.1 \%$
&   \\
$r_{\rm r.m.s.}^{\rm G}({}^{3}{\rm He}^{1/2+})$ & [fm] & 
      1.86  & (8.98\%) %$9.55 \%$ 
&   & 1.76  & (6.96\%) %$14.1 \%$ 
&   \\
$r_{\rm r.m.s.}^{\rm max}({}^{3}{\rm He}^{1/2+})$ & [fm]  & 
      1.88  & (9.00\%) %$8.54 \%$ 
&   & 1.78  & (7.05\%) %$13.1 \%$ 
&  \\
\cline{3-8}
$r_{\rm r.m.s.}^{\rm exp}({}^{3}{\rm He}^{1/2+})$ & [fm] & 
\multicolumn{5}{|c}{1.97} &\cite{Til87}\\ 
\hline\hline
\caption{
  The charge ${\rm r.m.s.}$-radii $r_{\rm r.m.s.}$ 
  of the used model-spaces for the fragments in the ${}^4{\rm Li}$-scattering-system 
  are summarized.
  For comparison the experimental data (exp) and the best result of 
  a RRGM calculation (max) are shown for the different studied
  $NN$-interactions Argonne-$v_{14}$ and Bonn. The $D$-state probability
  of each model is shown in parentheses.
  }\label{tab:rms}
\end{longtable}
\end{table}

%%%%%%%%%%%%%%%%%%%%%%%%%%%%%%%%%%%%%%%%%%%%%%%%%%%%%%%

\section{Results and Discussion}\label{sec:results}

\subsection{$(p-{}^3{\rm He})$ phase-shifts}

In this subsection phase-shifts are diagonal-phase-shifts. Published
eigen-phase-shifts have been transformed to diagonal-phase-shifts.

Unfortunately there are no modern theoretical phase-shift calculations 
at higher energies up to $E_{\rm c.m.}$ $=$ $30$ MeV. 
Only Reichstein et al. \cite{Rei71} 
and Heiss et al. \cite{Hei72} did Resonating Group Model 
calculations using a much smaller model-space 
and a much simpler $NN$-interaction in the considered energy range.   
Reichstein et al.
used a purely central $NN$-potential. Heiss et al. 
used a softcore central, spin-orbit and tensor-force potential.
Beltramin et al. \cite{Bel85} made a separable potential model up to 
$E_{\rm c.m.}$ $=$ $10$ MeV. 
We will not compare their results with ours since their models are
much simpler than ours.
Viviani et al. \cite{Viv01} recently studied the elastic
$(p-{}^3\mbox{He})$-scattering-system 
using the Kohn variational principle with correlated hyperspherical 
harmonics. As interaction  Viviani et al. used 
the $av_{18}$-$NN$-interaction \cite{Wir95} and the 
$av_{18}$-$NN$-interaction with the  Urbana IX $3N$-potential 
\cite{Pud97}. Unfortunately they published only results 
for very low energies ($E_{\rm c.m.}\leq 1.69$ MeV).
Pfitzinger et al. \cite{Pfi01} did an $R$-matrix analysis and an RRGM calculation
but only at low energies ($E_{\rm c.m.}<5$ MeV). 
We will not include the results of Pfitzinger et al. into the figures. 
In the following we will compare our results with the phase-shift analysis
based on experimental data in the desired energy-region 
of Darves-Blanc et al. \cite{Dar72}, McSherry et al. \cite{McS70},
M\"uller et al. \cite{Mul78} and Tombrello \cite{Tom64}.

\begin{figure}[ht]
  \begin{center}
    \epsfig{file=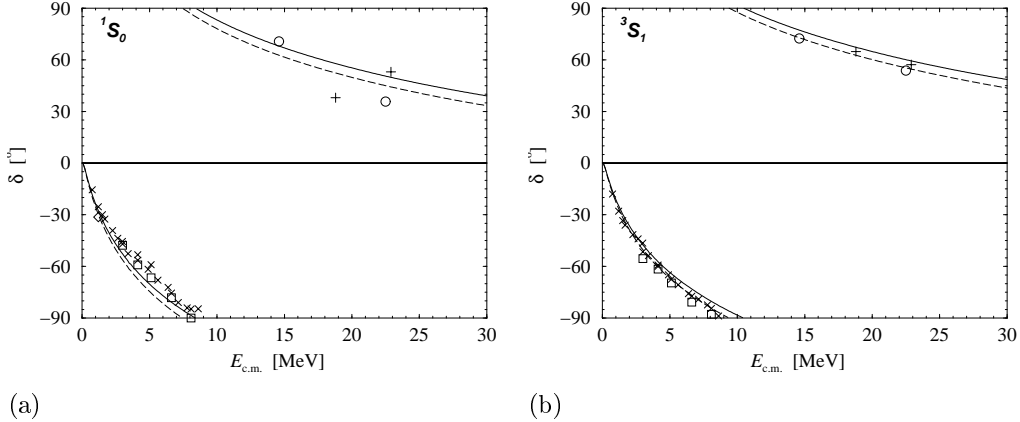}
    \caption{\label{fig:ps1S03S1}
      The ${}^1S_0$ (a) and the ${}^3S_1$ phase-shifts (b) of 
      $(p-{}^3{\rm He})$ scattering using the medium model-space for
      the ${}^3{\rm He}$ wave-function for the Argonne-$v_{14}$ (dashed line) and the
      Bonn (solid line) interaction in comparison with experimental 
      and theoretical results from other groups is shown:
      Darves-Blanc et al. \cite{Dar72} ($\circ$), McSherry et al. \cite{McS70} ($\square$),
      M\"uller et al. \cite{Mul78} ($+$), Tombrello \cite{Tom64} ($\times$)
      and Viviani et al. \cite{Viv01} ($\diamond$).
      }
  \end{center}
\end{figure}

\begin{figure}[ht]         
  \begin{center}
    \epsfig{file=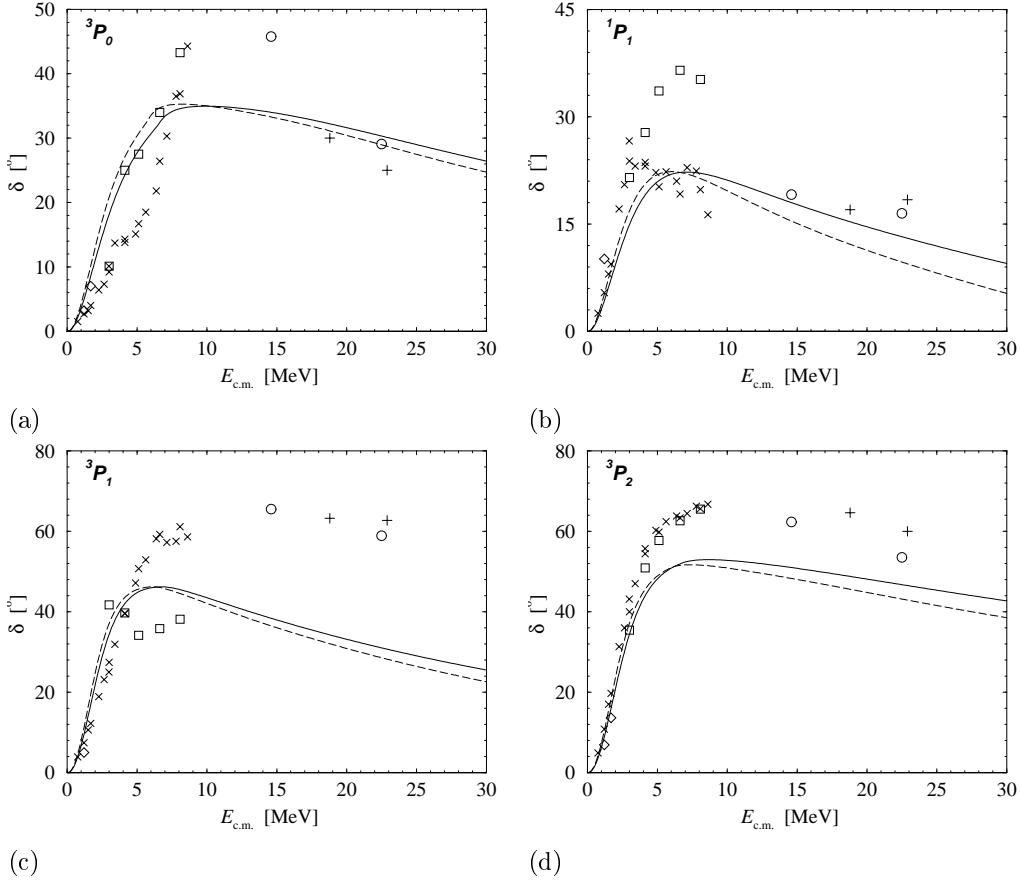}
  \caption{\label{fig:ps3P01P13P13P2}
    The same as in figure \ref{fig:ps1S03S1} is shown but for
    the ${}^3P_0$ (a), ${}^1P_1$ (b), ${}^3P_1$ (c) and ${}^3P_2$ (d)
    phase-shifts.
    }
  \end{center}
\end{figure}

\begin{figure}[ht]
  \epsfig{file=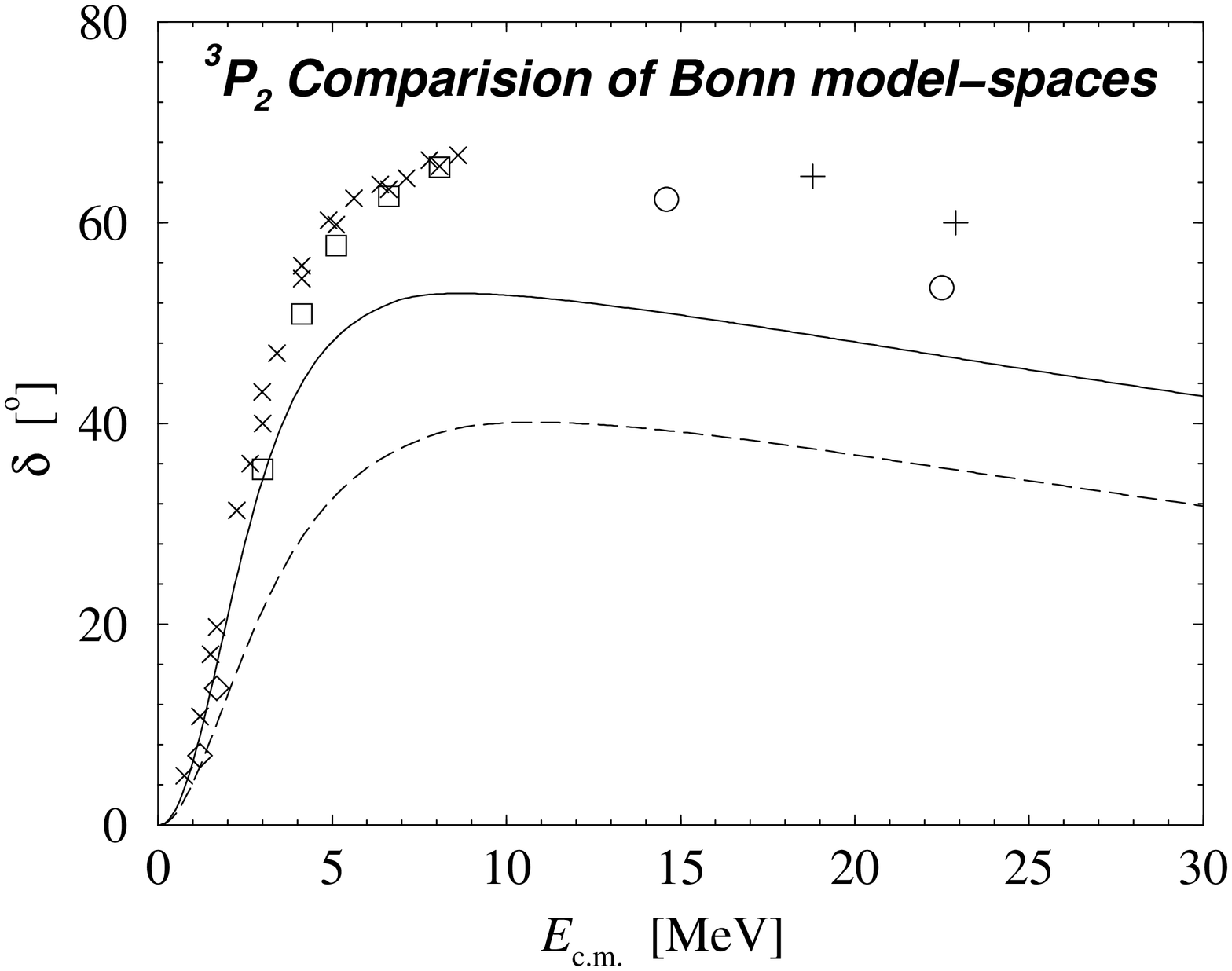, width=12.6cm, clip=, angle=-90}
  \caption{\label{fig:3p2}
    The ${}^3P_2$ phase-shift is shown using the
    Bonn interaction for the medium (solid line) and the large model (dashed line) 
    for the ${}^3{\rm He}$ wave-function. For comparison results 
    of other experimental and theoretical groups are shown as indicated 
    in figure \ref{fig:ps1S03S1}.
    } 
\end{figure}

%%%%%%%%%%%%%%%%%%%
% mittleres System
%
The phase-shifts obtained by the RRGM calculation using the 
medium model-space are compared with the results of other groups in 
Fig. \ref{fig:ps1S03S1}-\ref{fig:3p2}. 
The $S$-wave phase-shifts, shown in figure \ref{fig:ps1S03S1}, 
are in very good agreement in the whole range of interest with the 
results obtained by others \cite{Viv01,Dar72,McS70,Mul78,Tom64}.
The $S$-wave phase-shifts are all negative due to the underlying
Pauli-forbidden states. 
The $D$- and $F$-wave phase-shifts are not shown
because they are small and they are in good agreement
with the results of others cited
where the $D$-wave phase-shifts are between $-10^\circ$ and $0^\circ$
in the whole energy range.
The $F$-wave phase-shifts are even smaller 
than the $D$-wave phase-shifts by a factor of about 10. 
For the $P$-wave phase-shifts, see figure \ref{fig:ps3P01P13P13P2},
however,
the situation is  different. They are all positive. 
The RRGM reproduces the over-all behavior quite well,
but the maximal values are never reached, that means that all $P$-wave 
phase-shifts are not attractive enough. Furthermore the maxima occur at too low energies. 
The Bonn-interaction reproduces these phase-shifts slightly better.
Since the Bonn-potential underbinds the ${}^3{\rm He}$ less than the 
Argonne-$v_{14}$-interaction (see Tab. \ref{tab:binding}). 
Both forces predict also a too small r.m.s.-radius in the studied model-spaces where 
Argonne-interaction gives a 
better description (see Tab. \ref{tab:rms}).
Generally the Bonn-interaction yields a slightly better description of the  
${}^3{\rm He}$ subsystem than the Argonne-$v_{14}$-force 
and yields therefore to a slightly better description 
(see Tab. \ref{tab:binding} and Tab. \ref{tab:rms}).

The analysis of McSherry et al. and Tombrello disagree around $5$ MeV 
for the ${}^1P_1$ (Fig. \ref{fig:ps3P01P13P13P2} (b))
and ${}^3P_1$ (Fig. \ref{fig:ps3P01P13P13P2} (c)). 
From general consideration we consider the results of
Tombrello more reliable, as they indicate only a weak j-splitting. Our 
results for the $1^-$ phase-shifts are quite similar to those of 
Pfitzinger et al. \cite{Pfi01}.

But for the ${}^3P_2$ phase-shift analysis (Fig. \ref{fig:ps3P01P13P13P2} (d))
the results of the other groups are consistent. 
Here we have a good reproduction at low energies $\le 3$ MeV. Then
our phase-shift shows clearly not enough attraction.
Also Pfitzinger et al. \cite{Pfi01} reported that they are
not able the to reproduce the  ${}^3P_0$ and the  ${}^3P_2$
phase-shift.  Therefore we will concentrate  in the following  
on the ${}^3P_2$ phase-shift.

In Fig. \ref{fig:3p2} we compare the ${}^3P_2$ phase-shift 
calculated using the Bonn-interaction for
the medium and the large model-space with the results
of others as indicated. It turns out 
that the better ${}^{3}$He wavefunction does not necessarily
lead to a better description of the phase-shift 
although this model reproduces more accurately the binding-energy and
r.m.s.-radius (see Tab. \ref{tab:binding} and Tab. \ref{tab:rms}).
Similar results have been obtained using  the Argonne-$v_{14}$-interaction.
The overall description of the phase-shifts is improved using
the large model-space but the $P$-wave phase-shifts show less
attraction than in the medium model. 
In the medium model the 
$P$-wave phase-shifts drop too fast at higher energies because
of the too small r.m.s.-radius in comparison to the larger model-space.

%%%%%%%%%%%%%%%%%%%%%%%%%%%%%%%%%%%%%%%%%%%%%%%%%%%%%%%%%%%%%%%%%%%%%%%%%

\subsection{Analyzing powers $A_y$
and differential cross-section $\rmd\sigma/\rmd\Omega_{\rm c.m.}$}

To check if the ${}^3P_2$ phase-shift carries the main
contribution of the so-called $A_y$-problem we will take now a closer look
at analyzing-power $A_y$ and the differential cross-section
$\rmd\sigma/\rmd\Omega_{\rm c.m.}$.
We compare the analyzing-power $A_y$ and the differential cross-section at  
$E_{\rm c.m.}=5.20$ MeV and at $E_{\rm  c.m.}=22.5$ MeV with data, 
namely Birchall et al. \cite{Bir84}, McCamis et al. \cite{McC85}, 
McDonald et al. \cite{McD64} and  McSherry et al. \cite{McS70} for the 
analyzing-power results and McDonald et al. \cite{McD64} and
Harbison et al. \cite{Har70} for the cross-section results.
The error bars are increased in comparison to the original ones
of  McCamis et al. \cite{McC85} because we took the values from the their
figures since the data was not tabulated.

\begin{figure}[ht]
  \epsfig{file=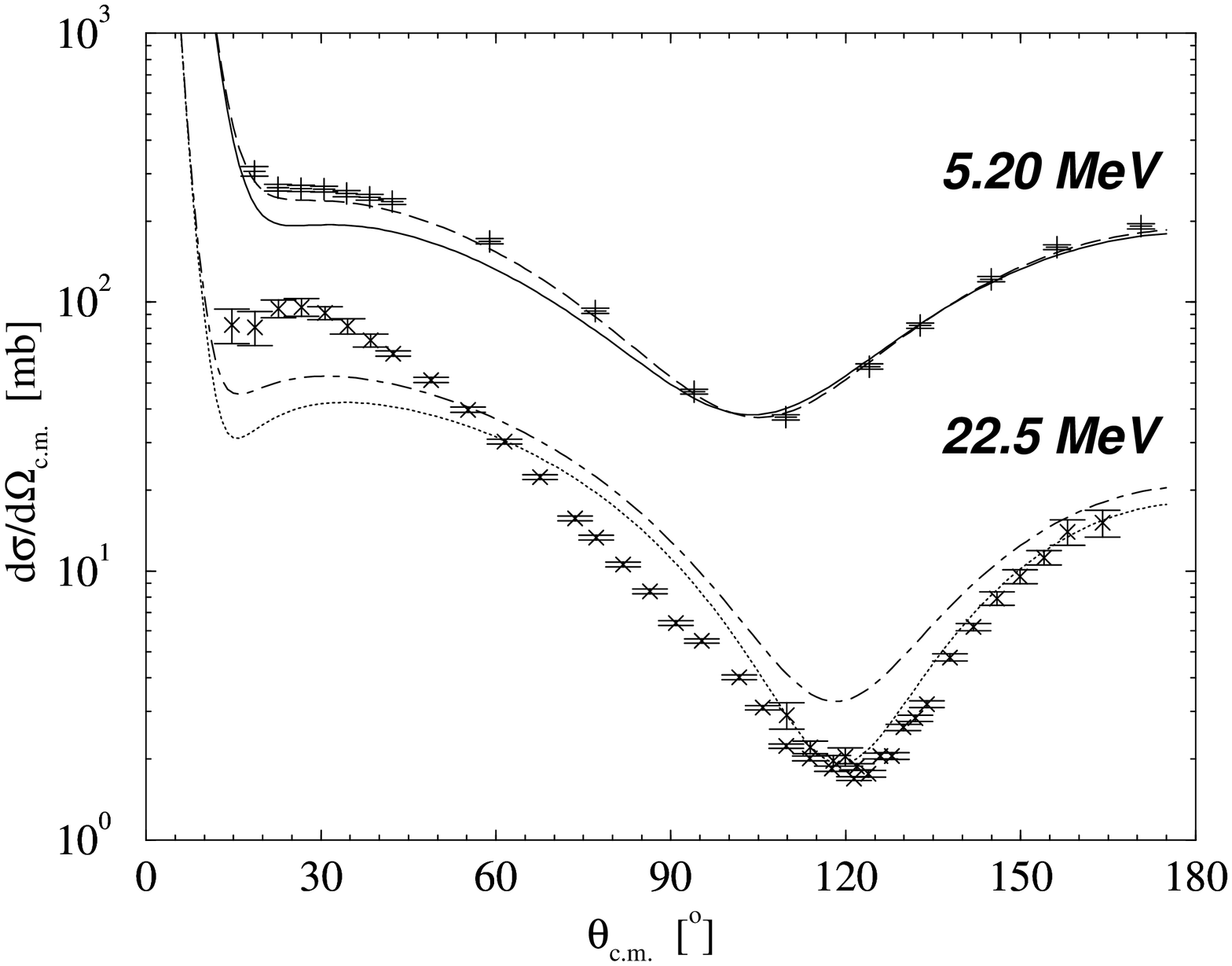, width=12.6cm, angle=-90}
  \caption{\label{fig:cs} 
    The differential cross-section $\rmd\sigma/\rmd\Omega_{\rm c.m.}$
    in comparison with the experiments
    at  $E_{\rm c.m.}=5.20$ MeV (solid line) and $E_{\rm c.m.}=22.5$ MeV 
    (dotted dashed line) is shown. 
    The curve where to the ${}^3P_2$ phase-shift $10.5^\circ$
    have be added at $E_{\rm c.m.}=5.20$ MeV is the dashed line.
    And the curve where to the ${}^3P_2$ phase-shift  $15.0^\circ$
    have been added at $E_{\rm c.m.}=22.5$ MeV is the  dotted  line.
    The experimental data at $E_{\rm c.m.}=5.21$ MeV is taken form
    McDonald et al. \cite{McD64} and at  $E_{\rm c.m.}=23.0$ MeV from
    Harbison et al. \cite{Har70}.
    } 
\end{figure}

In Fig. \ref{fig:cs}  we compare the results for the differential cross-section 
of McDonald et al. 
at  $E_{\rm c.m.}=5.21$ MeV
and Harbison et al. 
at  $E_{\rm c.m.}=23.0$ MeV with our calculation in
the medium Bonn-model at $E_{\rm c.m.}=5.20$ MeV (solid line) 
and $E_{\rm c.m.}=22.5$ MeV (dotted dashed line).
The curves in Fig. \ref{fig:cs} (dashed line and dotted line)
correspond to calculations where to the ${}^{3}P_{2}$ phase-shift has be modified.
We have added 
at $E_{\rm c.m.}=5.20$ MeV $10.5^\circ$ and at $E_{\rm c.m.}=22.5$ MeV $15.0^\circ$
to reach the experimental ${}^{3}P_{2}$ phase-shift
in the $S$-matrix-element. Figure \ref{fig:cs}  shows that we are now in very
good agreement with the experiment at $E_{\rm c.m.}=5.20$ MeV.
But at $E_{\rm c.m.}=22.5$ MeV it turns out that a better
description of the ${}^{3}P_{2}$ phase-shift is not enough to explain
the experimental data. At least we get a good agreement for angles
above $110^\circ$.  

\begin{figure}[ht]
  \begin{center}
    \epsfig{file=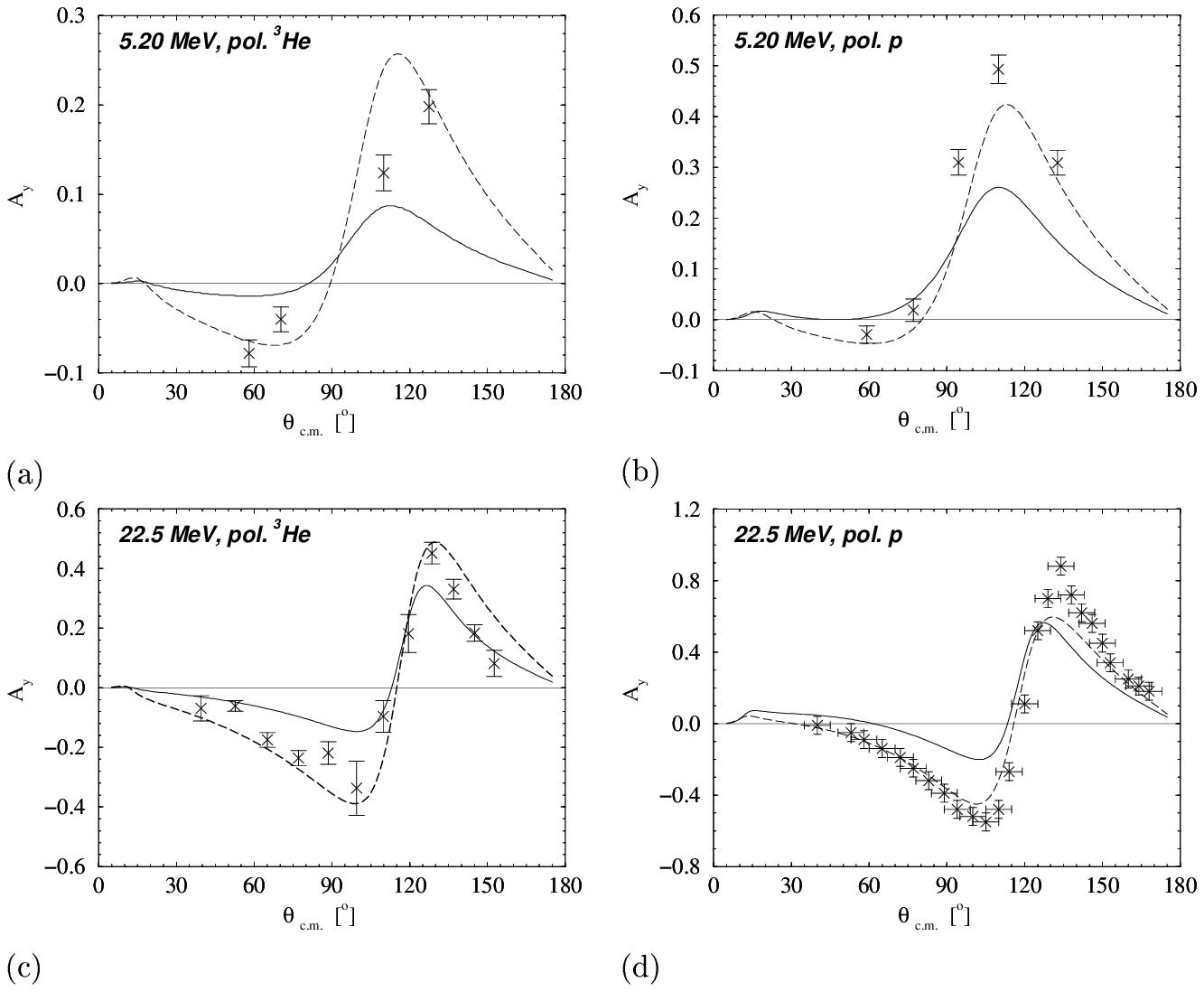}
    \caption{\label{fig:ay} 
      The analyzing power $A_y$ (solid line) is shown in
      comparison with experimental results ($\times$)
      with polarized ${}^3{\rm He}$ at $E_{\rm c.m.}=5.20$ MeV  and
      results of McSherry et al. \cite{McS70} at $E_{\rm c.m.}=5.18$ MeV (a),   
      with polarized protons at $E_{\rm c.m.}=5.20$ MeV  and 
      results of McDonald et al. \cite{McD64} at $E_{\rm c.m.}=5.21$ MeV (b), 
      with polarized ${}^3{\rm He}$ at $E_{\rm c.m.}=22.5$ MeV and 
      results of McCamis et al. \cite{McC85} (c)
      and
      with polarized protons at $E_{\rm c.m.}=22.5$ MeV and 
      results of Birchall et al. \cite{Bir84} (d).  
      The curve where to the ${}^3P_2$ phase-shift 
      $10.5^\circ$ (subfigure (a) and (b)) and $15.0^\circ$ (subfigure (c) and (d))
      have been added to reach the experimental results
      is the dashed line.
      }
  \end{center}
\end{figure}

Since looking on the differential cross-section can only provide
a rough overview of the quality of the calculation because the differential cross-section
is not too sensitive to the particle spin.  
Therefore we study now to the much more sensitive analyzing-powers 
(see Fig. \ref{fig:ay}). 
We employed the same modifications as for the  differential cross-section, 
setting the ${}^{3}P_{2}$ phase-shift
to its experimental value. 
The calculated analyzing-powers $A_y$ demonstrate  that  ${}^{3}P_{2}$ phase-shift
has the main contribution to the $A_y$-problem. In fact all $P$-wave phase-shifts 
show too less attraction. 
Including the ${}^{3}F_{2}$-channel does not the change the results. 
Therefore the tensor-force has no strong contribution to the $A_y$ problem.

%##############################################################################
% CONCLUSIONS
%##############################################################################

\subsection{Conclusions}

We have discussed a new RRGM calculation of the ${}^4{\rm Li}$-scattering
system using the Argonne-$v_{14}$- and the Bonn-interaction and two
different model-spaces. The calculated phase-shifts have been compared to
experimentally  and theoretical obtained ones. The overall description
of the phase-shifts is in very good agreement with the results of others
\cite{Viv01,Dar72,McS70,Mul78,Tom64}. 
Due to the fact that the Bonn-interaction is slightly more attractive
than the Argonne-$v_{14}$-interaction leads to a better reproduction 
of the experimental data by the Bonn-potential. 
But the $P$-wave phase-shifts show not enough  attraction. This
result was also obtained by Pfitzinger et al. \cite{Pfi01} and Viviani
et al. \cite{Viv01}. Pfitzinger et al. and Viviani et al. have used the Argonne-$v_{18}$-interaction
with three-particle-forces. At low energies the analyzing-power $A_y$
can be reproduced by setting the ${}^3P_2$ phase-shift to the experimental
determined one.  Pfitzinger et al. \cite{Pfi01} pointed out that also
the ${}^3P_0$ phase-shift has to be corrected to get into agreement.
Unfortunately our model-space is too small to make a statement to
this result. But at higher energies it turns clearly out
that not only the ${}^3P_2$ phase-shift has to be corrected to reproduce
analyzing-powers and cross-sections. And at low energies the contribution
of three-particle-forces is negligible since we can reproduce 
all available phase-shifts within our model. The $NN$-interactions used yield large
differences in the $P$-wave phase-shifts compared to experimental values. 
Pfitzinger et al. \cite{Pfi01} have demonstrated that new contributions 
to the 3$N$-forces acting on  $P$-waves should be considered especially for the
description of the ${}^3P_2-{}^3P_0$ splitting.  
Therefore we consider this system well suited for further studies of the $3N$-forces.
All the realistic $NN$-interactions studied till
now, $av_8'$, $av_{14}$, $av_{18}$, and Bonn show the same behavior for all the $P$-wave phase
shifts.

%##############################################################################
% Acknowledgments
%##############################################################################

\section*{Acknowledgments}

The authors like to thank 
C. Winkler and J. Wurzer for fruitful discussions
during this work. And in addition B. Pfitzinger not only for discussions
but also for providing us with model-spaces for the fragments of
the ${}^4{\rm Li}$-scattering-system. C.R. likes
to thank W. Leidemann 
for discussions of the 
phase-shifts 
in the $(n-{}^3{\rm He})$ scattering system.

%##############################################################################
% References
%##############################################################################

\end{document}